\documentclass[pra,preprint]{revtex4}
\usepackage{epsfig}
\usepackage{graphicx}
\usepackage{epstopdf}
\usepackage{amsmath}

\newcommand{\be}{\begin{equation}}
\newcommand{\ee}{\end{equation}}
\newcommand{\bea}{\begin{eqnarray}}
\newcommand{\eea}{\end{eqnarray}}
\newcommand{\vv}[1]{\mbox{\bf #1}}

\begin{document}
\title{A semiclassical hybrid approach to linear response functions for infrared spectroscopy}
\author{Frank Grossmann}
\affiliation{Institute for Theoretical Physics, 
\\
Technische Universit\"at Dresden,
\\
D-01062 Dresden, Germany}
\date{\today}

\begin{abstract}
Based on the integral representation of the semiclassical propagator of Herman and Kluk (HK),
and in the limit of high temperatures, we formulate a hybrid expression for the correlation function
of infrared spectroscopy. This is achieved by performing a partial linearization inside the
integral over the difference of phase space variables that occurs after a twofold application of the
HK propagator. A numerical case study for a coupled anharmonic oscillator shows that already for a total number of only two
degrees of freedom, one of which is treated in the simplified manner, a substantial reduction of the
numerical effort is achieved.
\end{abstract}  

\maketitle

\newpage
\section{Introduction}
The calculation of quantum mechanical correlation functions plays a central role in the theoretical understanding
of the interaction between matter and radiation or particles \cite{Tan06}. In the case of (infrared) IR spectroscopy
the relevant linear response correlation function is given by \cite{Muka}
\be
R^{(1)}(t)=\frac{i}{\hbar}{\rm Tr}\{\hat{\vv q}^{\rm T}(t)[\hat{\vv q},\hat{\rho}]\},
\label{eq:R1q}
\ee
where $\hat{\vv q}$ is denoting the position space operator of the infrared active degrees of freedom (DOFs)
(which may be a subset of the total number of DOFs) and the proportionality constant, relating position to 
the dipole operator is suppressed. $\hat\rho$ is the equilibrium thermal density operator of the complete system.
The calculation of the expression above, e. g., in position or energy basis becomes more and more evolved
the higher the total number of degrees of freedom. Approximative ways to calculate the correlation
function are therefore highly desirable.

One way to proceed is to approximate the density matrix by its high temperature limit and to approximate the
two time-evolution operators appearing in the Heisenberg operator $\hat{\vv q}(t)$
by semiclassical expressions \cite{NEL03} of Herman-Kluk (HK) type \cite{HK84}. Numerically even this approximate approach is barely
possible if the number of DOFs exceeds, say 3 or 4, because of the emergence of two semiclassical propagators, 
for every time step, double phase space integrals have to be done, which, for 4 DOFs amounts already to a 16 fold 
integral to be calculated at each time step.
A formidable reduction of the numerical effort can be achieved by using classical mechanics, e.g., in a classical
Wigner (i.\ e.\ linearized semiclassical) description. There the double phase space integral is reduced to a single
phase space integral after performing the integral over the difference of phase space variables in
a linearization approximation. Quantum effects, like beatings in the time-signal due to the decrease of the anharmonic oscillator's level spacing 
thereby are lost, however \cite{GL08,jcp14}. This shortcoming can be circumvented by refraining from the
use of purely classical trajectories. In \cite{LM07}, four different ways of improving on the purely classical
approach but sticking to the single phase space integral are compared. Alternatively, the mean trajectory approach of
Loring and collaborators is still using classical trajectories but incorporates an additional quantization condition
of an action into the LSC-IVR expression \cite{GL08,MAL15}.

Here we want to follow a different approach that is close in spirit to the semiclassical hydrid
approach to many particle quantum dynamics put forth previously \cite{jcp06}. We keep the full complexity of the double phase space
integral in the DOF that is infrared active and perform a linearization leading to an LSC-IVR-type expression
in the remaining degrees of freedom. This way, we gain a result whose complexity is dramatically
reduced if the number of inactive degrees of freedom is large. Retaining the full (semiclassical) complexity for the
IR active degree of freedom will allow, however, to still describe some relevant quantum features
in the dynamics.

The paper is organized as follows: In Section \ref{sec:SHIR} the full HK approach to the linear response function
is briefly reviewed. Then the hybrid approach to that quantity is introduced, which is based on a partial
linearization of the full semiclassical expression. Numerical results at different levels of approximation 
are then shown in Section \ref{sec:num} for a Morse oscillator coupled a harmonic bath degree of freedom.
Finally some conclusions and an outlook are given. In the Appendix a rederivation of the fully linearized
version of the semiclassical theory is given.

\section{Semiclassical hybrid expression for IR spectroscopy}
\label{sec:SHIR}

As shown in \cite{NEL03,GL08} the semiclassical linear response (or correlation) function for IR spectroscopy
for a system of $N$ degrees of freedom, in the case of high temperature, is given by
\bea
R^{(1)}_{\rm HK}(t)&=&\frac{\beta}{m Q}\frac{1}{(2\pi\hbar)^{2N}}\int {\rm d}^{2N}z_1\int {\rm d}^{2N}z_2
\exp\{-\beta H(\bar{\vv z})\}\bar{\vv p}^{\rm T}(\bar{\vv q}_t+\frac{i}{2\hbar\gamma}\Delta \vv p_t)
\nonumber
\\
\label{eq:R1}
&&C(\vv z_1,t)C^\ast(\vv z_2,t)\langle\vv z_{2,t}|\vv z_{1,t}\rangle\langle\vv z_1|\vv z_2\rangle
\exp\left\{\frac{i}{\hbar}[S(\vv z_1,t)-S(\vv z_2,t)]\right\},
\eea
where the (row) vectors $\vv z^{\rm T}_i=(\vv p^{\rm T}_i,\vv q^{\rm T}_i)$, with $i=1,2$, consist of momenta and position vectors 
and 
\bea
\bar{\vv z}&=&\frac{\vv z_1+\vv z_2}{2}
\label{eq:sum}
\\
\Delta \vv z&=&\vv z_1-\vv z_2
\label{eq:dif}
\eea
denote the average and difference vectors, respectively, and the $|\vv z_i\rangle$ indicate ket vectors that, in position representation, are 
normalized Gaussians
\be
\langle \vv x|\vv z_i\rangle=\left(\frac{\det{\bf \gamma}}{\pi^N}\right)^{1/4}
\exp\left\{-\frac{1}{2}(\vv x-\vv q_i)^{\rm T}{\bf \gamma}(\vv x-\vv q_i)+
\frac{i}{\hbar} \vv p^{\rm T}_i(\vv x-\vv q_i)\right\}.
\ee
Furthermore, use has been made of the helpful identities \cite{NEL03}
\bea
\langle \vv z_1|[\hat{\vv q},\hat{A}]|\vv z_2\rangle&=&i\hbar\langle \vv z_1|\vv z_2\rangle\frac{\partial}{\partial \bar{\vv p}}
\frac{\langle \vv z_1|\hat{A}|\vv z_2\rangle}{\langle \vv z_1|\vv z_2\rangle}
\\
\frac{\langle \vv z_1|\hat{\rho}|\vv z_2\rangle}{\langle \vv z_1|\vv z_2\rangle}&=&\frac{\exp\{-\beta H(\bar{\vv z})\}}{Q},
\eea
the second of which being valid in the high temperature limit for a normalized density matrix $\hat{\rho}$ with $H(\bar{\vv z})$ the classical
Hamiltonian (taken at the average variables) and $Q$ the (classical) partition function, see also \cite{HC99}. The high temperature limit
has been seen to yield surprisingly good results, also for intermediate temperatures, see e.\ g.\ \cite{NEL03,jcp14}. In passing we note
that the full integral expression in (\ref{eq:R1}) is real (as it has to be by comparison to (\ref{eq:R1q})), as can be seen by the structure 
of the integrand.

The dynamical part of the approximate response function has been gained by the use of the semiclassical Herman-Kluk (HK) \cite{HK84} 
time-evolution operator
\bea
\label{eq:HK1}
\exp\{-i\hat{H}t/\hbar\}\approx
\int\frac{{\rm d}^{2N}z}{(2\pi\hbar)^N}
C(\vv z,t)|\vv z_t\rangle
\exp\left\{\frac{i}{\hbar}S(\vv z,t)\right\}
\langle \vv z|
\eea
with the HK prefactor 
\be
C(\vv z,t)=\sqrt{\det {\bf h}},
\ee
where
\be
\label{eq:pre}
{\bf h}=
\frac{1}{2}({\bf m}_{11}+\gamma{\bf m}_{22}\gamma^{-1}
-i\hbar\gamma{\bf m}_{21}-\frac{1}{i\hbar}{\bf m}_{12}\gamma^{-1})
\ee
for diagonal (not necessarily proportional to the unit matrix) width-parameter matrix $\gamma$ 
with real and positive elements. The ${\bf m}_{ij}$ are sub-blocks of the monodromy matrix
\be
{\bf M}\equiv
\left(\begin{array}{cc}
{\bf m}_{11}&{\bf m}_{12}\\
{\bf m}_{21}&{\bf m}_{22}
\end{array}\right)
\equiv
\left(\begin{array}{cc}
\frac{\partial\vv p_{t}}{\partial\vv p'^{\rm T}}&
\frac{\partial\vv p_{t}}{\partial\vv q'^{\rm T}}\\
\frac{\partial\vv q_{t}}{\partial\vv p'^{\rm T}}&
\frac{\partial\vv q_{t}}{\partial\vv q'^{\rm T}}
\end{array}\right).
\ee
Furthermore, $S(\vv z,t)=\int_0^t {\rm d}t'\, L$ is the classical action functional with the Lagrangian $L=T-V$.
Frequently, the HK propagator is referred to as a semiclassical initial value representation (SC-IVR) of the propagator, because the only dynamical
quantities that enter the final expression are solutions of classical initial value problems. Its historical precursor is the
frozen Gaussian wavepacket dynamics of Heller \cite{He81}.

Another prominent initial value representation, but this time of the wavefunction, is the thawed Gaussian wavepacket 
dynamics (TGWD) of Heller \cite{He75}, which is based on a single classical trajectory (the center trajectory of the
Gaussian wavepacket). There is a close connection between the Herman-Kluk propagator applied to a Gaussian
wavepacket and TGWD. By doing an expansion of the exponent in the HK expression
up to second order around the wavepacket center (also referred to as ``linearization'', because 
the positions and momenta are expanded up to first order) and performing the resulting Gaussian integral,
TGWD follows from the HK-propagator applied to a Gaussian wavepacket \cite{camp99,DE06,jcp06}. We stress that this procedure
is inferior to a stationary phase approximation; the TGWD therefore is 
not a strict semiclassical theory. The question, however, is if something similar can be
done for the correlation function of IR spectroscopy. As noted previously \cite{NEL03} and as shown in the appendix \ref{app:stab}, 
using calculus analogous to the one used in \cite{jcp06}, this is indeed the case. There exists, however, no wavepacket center in the
semiclassical correlation function for IR spectroscopy to expand 
around, and the linearization is performed in the difference variables after a transformation to sum and
difference variables for the double phase space integral in (\ref{eq:R1}). The final result (\ref{eq:lscivr}) is
referred to as the linearized semiclassical initial value representation (LSC-IVR) for IR spectroscopy \cite{NEL03,SG03,LM07}.

This connection between the full HK and the linearized expression now serves as the starting point
to the hybrid approach to IR spectroscopy. First we rewrite Eq. (\ref{eq:R1}) in a form appropriate for numerical
calculations \cite{NEL03}
\bea
R^{(1)}_{\rm HK}(t)&=&\frac{\beta}{m Q}\frac{1}{(2\pi\hbar)^{2N}}\int {\rm d}^{2N}\bar{z}\exp\{-\beta H(\bar{\vv z})\}\bar{\vv p}^{\rm T}
\int {\rm d}^{2N}\Delta z(\bar{\vv q}_t+\frac{i}{2\hbar\gamma}\Delta \vv p_t)
\nonumber
\\
\label{eq:R1num}
&&C(\bar{\vv z}+\Delta \vv z/2,t)C^\ast(\bar{\vv z}-\Delta \vv z/2,t)
\exp\left\{\frac{i}{\hbar}[S(\bar{\vv z}+\Delta \vv z/2,t)-S(\bar{\vv z}-\Delta \vv z/2,t)]\right\}
\nonumber
\\
&&\exp\Bigl\{-\frac{1}{4}(\Delta \vv q^{\rm T}_t\gamma  \Delta \vv q_t+\Delta \vv q^{\rm T}\gamma  \Delta \vv q)
-\frac{i}{\hbar}\bar{\vv p}_t^{\rm T}\Delta \vv q_t+\frac{i}{\hbar}\bar{\vv p}^{\rm T}\Delta \vv q
\nonumber
\\
&&-\frac{1}{4\hbar^2}(\Delta \vv p^{\rm T}_t \gamma^{-1}\Delta \vv p_t+\Delta \vv p^{\rm T} \gamma^{-1}\Delta \vv p)\Bigr\},
\eea
by integrating over sum and difference variables, defined in (\ref{eq:sum},\ref{eq:dif}).
Now we assume, that there are $m$ (anharmonic) IR active
modes of a molecule coupled to a number $n$ of, e. g., harmonic modes, that could either be modes of the same molecule
or could be modes of a solvent environment. Then we will keep the full HK expression for 
the anharmonic modes (our ``system of interest'') and will perform a transition to the classical (linearized)
form of the expression for the remaining degrees of freedom in a similar spirit as it was done for the wavefunction in \cite{jcp06}.

To proceed, we denote the momenta and coordinates of the portion of the
total number of degrees of freedom (DOF), which we want to treat with the full HK approach, i.\ e.,
the IR active DOF, by the ``system'' vectors $\vv p_S$ and $\vv q_S$ with $m=N-n$ entries. For the
remaining ``bath'' DOF we use the phase space variables $\vv p_B$ and $\vv q_B$ of ``dimension'' $n$. The double phase space integration over
the harmonic modes shall now be treated in a linearized fashion as indicated in the appendix. In this way that sub phase-space double-integral condenses
into a single phase-space integral and a hybrid expression of the form 
\bea
R^{(1)}_{\rm hy}(t)&=&\frac{\beta}{m Q}\frac{1}{(2\pi\hbar)^{N+m}}\int {\rm d}^{2N}\bar{z}\exp\{-\beta H(\bar{\vv z})\}\bar{\vv p}_S^{\rm T}
\int {\rm d}^{2m}\Delta z_S(\bar{\vv q}_{S,t}+\frac{i}{2\hbar\gamma}\Delta \vv p_{S,t})
\nonumber
\\
\label{eq:R1hyb}
&&\sqrt{\frac{|{\bf h}(\bar{\vv z}+\widetilde{\Delta \vv z}/2,t)||{\bf h}^\ast(\bar{\vv z}-\widetilde{\Delta \vv z}/2,t)|}{(4\hbar^2)^n|{\bf A}_B|}}
\exp\left\{\frac{i}{\hbar}[S(\bar{\vv z}+\widetilde{\Delta \vv z}/2,t)-S(\bar{\vv z}-\widetilde{\Delta \vv z}/2,t)]\right\}
\nonumber
\\
&&\exp\Bigl\{-\frac{1}{4}(\Delta \vv q^{\rm T}_{S,t} \gamma_S  \Delta \vv q_{S,t}
+\Delta \vv q^{\rm T}_S\gamma_S \Delta \vv q_S)
-\frac{i}{\hbar}\bar{\vv p}_{S,t}^{\rm T}\Delta \vv q_{S,t}+\frac{i}{\hbar}\bar{\vv p}^{\rm T}_S\Delta \vv q_S
\nonumber
\\
&&-\frac{1}{4\hbar^2}(\Delta \vv p^{\rm T}_{S,t} \gamma^{-1}_S\Delta \vv p_{S,t}
+\Delta \vv p^{\rm T}_S \gamma^{-1}_S\Delta \vv p_S)\Bigr\},
\eea
where $\widetilde{\Delta \vv z}$ denotes difference variables which are zero in the harmonic DOFs, $\gamma_S$
is the sub-block of the width parameter matrix corresponding to the system DOFs (the width parameter matrix contains no coupling between its subblocks), 
and the vertical bars under the square root denote taking the determinant.

Furthermore, we used the $2n\times 2n$ matrix
\be
{\bf A}_B=\frac{1}{4\hbar^2}
\left(\begin{array}{cc}
\gamma_{B}^{-1}+\tilde{\bf m}_{11}^{\rm T}\gamma^{-1} \tilde{\bf m}_{11}+\tilde{\bf m}_{21}^{\rm T}\hbar^2\gamma\tilde{\bf m}_{21}
&
\tilde{\bf m}_{21}^{\rm T}\hbar^2\gamma \tilde{\bf m}_{22}+\tilde{\bf m}_{11}^{\rm T}\gamma^{-1}\tilde{\bf m}_{12})
\\
\tilde{\bf m}_{22}^{\rm T}\hbar^2\gamma \tilde{\bf m}_{21}+\tilde{\bf m}_{12}^{\rm T}\gamma^{-1}\tilde{\bf m}_{11}
&
\hbar^2\gamma_{B} +\tilde{\bf m}_{22}^{\rm T}\hbar^2\gamma \tilde{\bf m}_{22}+\tilde{\bf m}_{12}^{\rm T}\gamma^{-1}\tilde{\bf m}_{12}
\end{array}\right),
\ee
with $\gamma_B$ denoting the bath sub-block of the width parameter matrix. The $\tilde{\bf m}_{ij}$ matrices
are rectangular $N\times n$ sub-blocks of the stability matrix, see also \cite{jcp06}.

Formally, the expression in Eq. (\ref{eq:R1hyb}) does not look as compact as the starting expression (\ref{eq:R1}) but for applications it has 
the decisive advantage to be a much less high-dimensional integral (the second phase space integral is only $2m$ dimensional), 
in complexity somewhere in between the full double phase space integral expression
and the linearized, single phase space integral (\ref{eq:lscivr}) of the appendix.
Similar ideas have appeared in the literature before. For related semiclassically spirited work, see \cite{SM971,OA98,AYA15}. 
Analogous simplifications of the double phase space integral may occur in the large body of work that is based on the forward-backward idea of the
Macri and Miller groups with or without using a Filinov transformation \cite{TM99,SM992,WTM00,TWM01}. It has been stressed, however, that
for dipole-dipole correlation functions, the standard forward-backward methods do not go beyond the level of LSC-IVR \cite{TWM01}.
A recent review of analogous quantum classical hybrid approaches is given in \cite{Kap15}.

\section{Numerical Results}
\label{sec:num}

The model system of interest that we study in the following is a 1D Morse oscillator with unit mass and the potential
\be
V_S(x_S)=D\left[1-\exp(-\alpha x_S)\right]^2.
\ee
The dimensionless potential parameters $D=100$ and  $\alpha=0.2\sqrt{2}$ are the same that have
been used in a dissipative case study based on hierarchal equations of motion \cite{jcp14}. The eigenenergies
(setting also $\hbar$ equal to unity)
\be
\label{eq:morseen}
E_n=\omega_{\rm e}(n+1/2)-x_{\rm e}\omega_{\rm e}(n+1/2)^2,\qquad n=0,1,\dots\;.
\ee
of the Morse potential \cite{Mor29} contain the two parameters $\omega_{\rm e}=\alpha\sqrt{2D}$ and $x_{\rm e}=\omega_{\rm e}/(4D)$,
corresponding to the frequency of harmonic oscillations around the potential minimum and the anharmonicity constant.

For this initial study of the method, and to be able to compare to exact quantum results, the number of bath degrees of freedom of unit mass
shall also be restricted to one, and the coupling between system and bath is taken bilinear such that the full Hamiltonian of
the 2 DOF problem is given by
\bea
\hat{H}&=&\hat{H}_S+\hat{H}_B+\hat{H}_{SB}
\nonumber
\\
&=&\frac{\hat{p}_S^2}{2}+V_S(x_S)+\left(\frac{\hat{p}_{B}^2}{2}+D(\chi\alpha x_B)^2-\gamma x_Sx_B\right),
\eea
where $\chi$ can be used to tune the harmonic mode in or out of ``resonance'' with the Morse oscillator.

\begin{figure}
 \includegraphics[width=12cm]{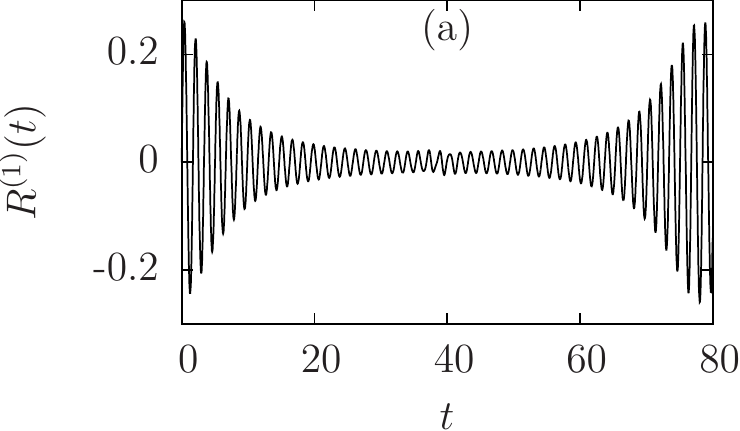}
 \includegraphics[width=12cm]{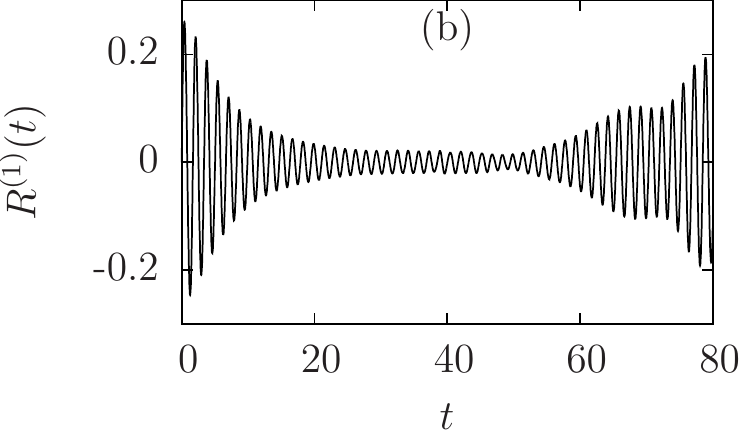}
 \caption{The fully quantum linear response correlation function of an infrared active 1D Morse oscillator 
 for dimensionless temperature $T=7$:
 (a) without coupling ($\gamma=0$)
 (b) with coupling ($\gamma=0.1$) to a harmonic oscillator with $\chi=0.9$}
 \label{fig:qu}
\end{figure}

\begin{figure}
 \includegraphics[width=12cm]{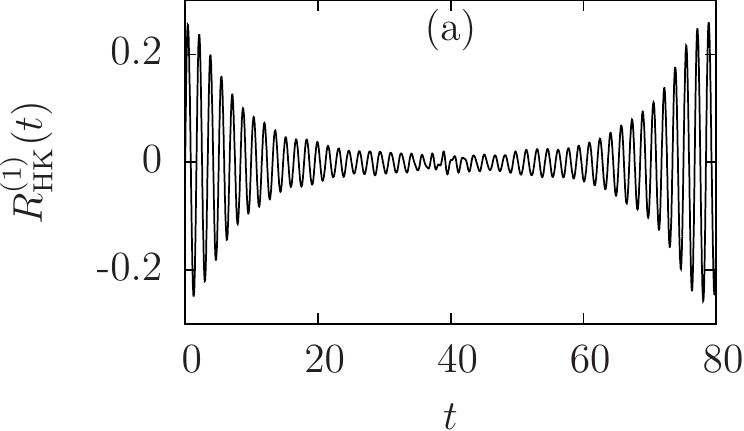}
 \includegraphics[width=12cm]{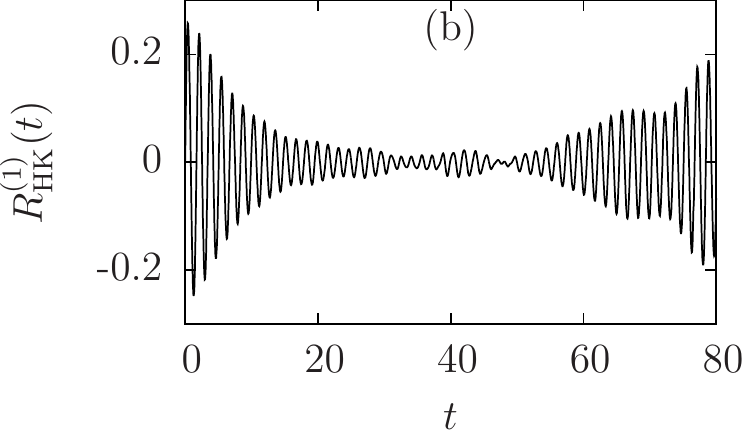}
 \caption{The HK linear response correlation function of an infrared active 1D Morse oscillator 
 for dimensionless temperature $T=7$:
 (a) without coupling ($\gamma=0$)
 (b) with coupling ($\gamma=0.1$) to a harmonic oscillator with $\chi=0.9$}
 \label{fig:sc}
\end{figure}

In the following, we will show comparisons of full quantum, full HK, full LSC-IVR and hybrid results for the
linear response function.  As has been highlighted in Fig.\ \ref{fig:qu} (a) of \cite{GL08}
as well as in \cite{jcp14} (see also Fig.\ 1 (a) below), in the case of just one single anharmonic degree of freedom without coupling to a bath mode,
the full quantum response function shows a beating pattern with fast
oscillations corresponding to roughly the harmonic frequency around the minimum of the potential curve
and recurrence periods proportional to $1/(\omega_{\rm e}x_{\rm e})$ \cite{GL08}. The full HK result, however, shows almost complete agreement 
with the full quantum result \cite{GL08} (see also Fig.\ \ref{fig:sc} (a) below).

For the potential parameters considered herein, we show a comparison between the uncoupled and the coupled correlation function
in the fully quantum case in Fig.\ \ref{fig:qu}.
There it can be seen that the coupling introduces additional complexity into the beating signal without
coupling, displayed in panel (a). So in panel (b) after a dimensionless time around $t=40$, the signal deviates
from the one in panel (a) and also the maximum recurrence of the signal at around $t=80$ is reduced.

Furthermore, for reasons of completeness, we also show the corresponding results in the LSC-IVR case of the Appendix
in Fig.\ \ref{fig:lsc}. Firstly, there is no recurrence in the signal without coupling to be observed.
As has been noticed in \cite{GL08} this recurrence is a quantum effect and the corresponding time
scale goes to infinity in the classical limit. The coupling induces additional complexity 
but the overall height of the time signal after around $t=30$ is marginal.
\begin{figure}
 \includegraphics[width=12cm]{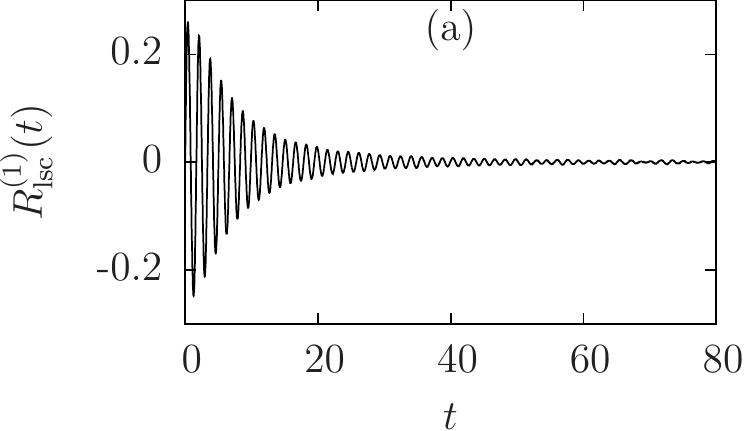}
 \includegraphics[width=12cm]{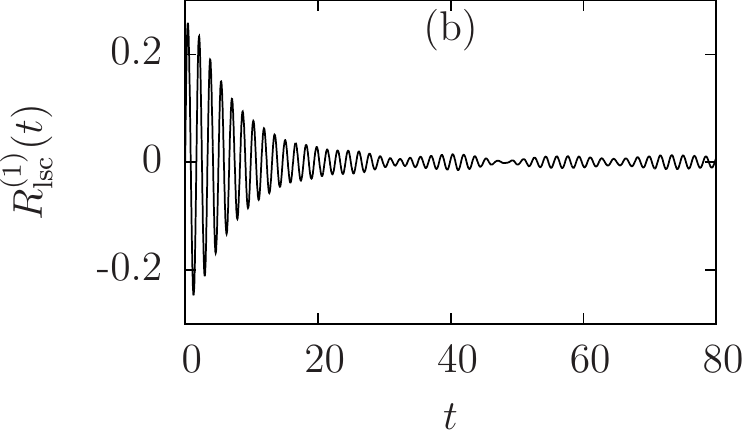}
 \caption{The LSC-IVR linear response correlation function of an infrared active 1D Morse oscillator 
 for dimensionless temperature $T=7$:
 (a) without coupling ($\gamma=0$)
 (b) with coupling ($\gamma=0.1$) to a harmonic oscillator with $\chi=0.9$}
 \label{fig:lsc}
\end{figure}

In Fig. \ref{fig:sc} the same comparison is now made between the corresponding full HK signals (both degrees of
freedom are sampled in the sum as well as in the difference phase space variables) and very similar (although not identical) 
results as in the quantum case of Fig. \ref{fig:qu} can be observed. 

\begin{figure}
 \includegraphics[width=12cm]{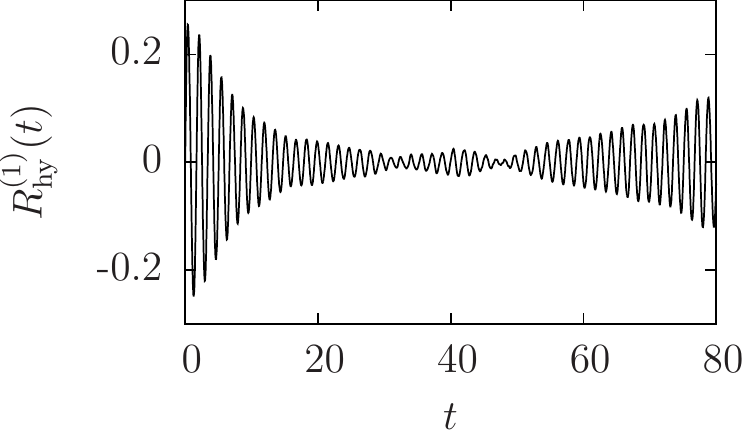}
 \caption{The hybrid linear response correlation function of an infrared active 1D Morse oscillator 
 for dimensionless temperature $T=7$ with coupling ($\gamma=0.1$) to a harmonic oscillator with $\chi=0.9$}
 \label{fig:hy}
\end{figure}

In Fig.\ \ref{fig:hy} the central result of the present study is displayed. This is an implementation of the correlation function 
of (\ref{eq:R1hyb}) in the hybrid case. It can be seen that, although the difference variables of the harmonic DOF is unsampled (i.\ e.\ it is described on the
linearized level of the appendix), the sampling of the difference variables of the anharmonic degree DOF is enough to reproduce the
recurrence in the time series to a surprisingly high degree.

The computational strategy to tackle the phase space integrals for the classical trajectory
based methods was to use Monte Carlo integration with importance sampling \cite{NUMREC} for both, the
integral over the sum as well as the one over the difference variables. Alternative strategies
can be envisaged (e.\ g.\ applying a Metropolis algorithm for the sum variable integral \cite{NEL03}). 
In the case of a single degree of freedom and for LSC-IVR 10$^6$ trajectories are enough for converged results
to within line thickness as long as the correlation function has not decayed to too small values and, interestingly, although a
double phase space integral has to be performed, 10$^6$ trajectories are also enough in the HK case for a single DOF. The HK calculations
require the determination of the stability information and the evaluation of exponentials and also of square roots of complex
numbers, however, and therefore are much more time consuming (by more than an order of magnitude) than the LSC-IVR ones.

The higher the dimensionality of the integral, the better the Monte Carlo method is suited. Therefore we do not
need a lot more trajectories to get converged results in the 2 DOF case for the LSC-IVR calculations. For the full
HK results that are plagued by a sign problem, however, an increasing number of trajectories is needed to get converged results for all times. We found that
under our sampling strategy 2$\times 10^7$ trajectories were necessary for convergence. This high number can be reduced by
more than an order of magnitude, if the hybrid idea put forth herein is used. In the hybrid case, whose results are displayed in Fig.\ \ref{fig:hy}
again only 10$^6$ trajectories were necessary for convergence in the 2 DOF case.

\section{Outlook and Conclusions}

We have formulated a semiclassical hybrid approach to linear IR spectroscopy which is based on the
same idea as the semiclassical hybrid dynamics put forth previously for wavefunctions \cite{jcp06}
as well as for density matrices \cite{jcp092}. A partial linearization
in the difference variables of the full double Herman Kluk expression for the correlation function leads to a 
working formula with a reduced dimensionality of the remaining integral to be performed numerically.
Already in the case of a single DOF treated in the simplified manner, a substantial reduction of the
numerical effort has been achieved, as demonstrated for an anharmonic oscillator coupled to a harmonic one.

In future work, we want to extend the idea presented here to arbitrary temperatures as well as to 
the calculation of higher order correlation functions needed for nonlinear spectroscopic techniques.

\acknowledgments
The author would like to thank Jiri Vanicek and Alfredo M.\ Ozorio de Almeida for valuable
discussions and the Max-Planck-Institute for the Physics of Complex Systems for the opportunity
to take part in the meetings of the Advanced Study group on ''Semiclassical Methods: Insight and Practice in 'Many' Dimensions''
led by Eric J. Heller and Steven Tomsovic.

Financial support by the Deutsche Forschungsgemeinschaft under GR 1210/4-2 is gratefully acknowledged.

\appendix

\section{LSC-IVR for $R^{(1)}(t)$ in the case of $N$ DOF}
\label{app:stab}

In this appendix we recall the derivation of the LSC-IVR approximation for linear IR spectroscopy. 
A simplification of the full HK expression (\ref{eq:R1}) can be achieved by expanding the
classical actions in the exponent around the mean phase space point $\bar{\vv z}$
up to second order, using
\bea
S(\vv z_1,t)&\approx&S(\bar{\vv z})+\frac{\partial S}{\partial \bar{\vv z}}(\vv z_1-\bar{\vv z})
+\frac{1}{2}\frac{\partial^2 S}{\partial \bar{\vv z}^2}(\vv z_1-\bar{\vv z})^2
\\
S(\vv z_2,t)&\approx&S(\bar{\vv z})+\frac{\partial S}{\partial \bar{\vv z}}(\vv z_2-\bar{\vv z})
+\frac{1}{2}\frac{\partial^2 S}{\partial \bar{\vv z}^2}(\vv z_2-\bar{\vv z})^2.
\eea
Taking the action difference, the second order term cancels and by using
\bea
\partial S/\partial \vv p&=&\vv p_t {\bf m}_{21}
\\
\partial S/\partial \vv q&=&\vv p_t {\bf m}_{22}-\vv p,
\eea
we get
\be
S(\vv z_1,t)-S(\vv z_2,t)\approx\bar{\vv p}_t {\bf m}_{21}\Delta \vv p+(\bar{\vv p}_t {\bf m}_{22}-\bar{\vv p})\Delta \vv q.
\ee
In the overlaps of the Gaussians appearing in the HK time-evolution operator
\bea
\label{eq:over}
\langle \vv z_{2,t}|\vv z_{1,t}\rangle&=&
\exp\Bigl\{-\frac{1}{4}\Delta \vv q^{\rm T}(t) \gamma  \Delta \vv q(t)
-\frac{i}{\hbar}\bar{\vv p}(t)^{\rm T}\Delta \vv q(t)
-\frac{1}{4\hbar^2}\Delta \vv p^{\rm T}(t) \gamma^{-1}\Delta \vv p(t)\Bigr\},
\eea
the  linear expansions
\bea
\label{eq:smat1}
\Delta \vv p(t)&=&{\bf m}_{11}\Delta \vv p+{\bf m}_{12}\Delta \vv q
\\
\label{eq:smat2}
\Delta \vv q(t)&=&{\bf m}_{21}\Delta \vv p+{\bf m}_{22}\Delta \vv q
\eea
are made and the phase factors in (\ref{eq:R1}) cancel out. Consistently, we take a zeroth order expansion
of the preexponential factor (i. e., we set $\Delta \vv z=0$ in the prefactor \cite{NEL03}) and change from
the volume elements ${\rm d}^{2N}z_1{\rm d}^{2N}z_2$ to ${\rm d}^{2N}\bar{z}{\rm d}^{2N}\Delta z$ (the absolute value of the 
Jacobian is unity). Then, the intermediate result
\begin{align}
\label{eq:R1inter}
R^{(1)}_{\rm lsc}(t)=\frac{\beta}{m Q}& \frac{1}{(2\pi\hbar)^{2N}}\int {\rm d}^{2N}\bar{z}\int {\rm d}^{2N}\Delta z
\exp\{-\beta H(\bar{\vv z})\}\bar{\vv p}^{\rm T}\bar{\vv q}(t)|C(\bar{\vv z},t)|^2
\nonumber
\\
& \exp\Bigl\{-\frac{1}{4\hbar^2}\Delta \vv p^{\rm T} (\gamma^{-1}
+{\bf m}_{11}^{\rm T}\gamma^{-1} {\bf m}_{11}
+\hbar^2{\bf m}_{21}^{\rm T}\gamma{\bf m}_{21}) \Delta \vv p
\nonumber
\\
& -\frac{1}{4}\Delta \vv q^{\rm T}(\gamma +{\bf m}_{22}^{\rm T}\gamma {\bf m}_{22}+\frac{1}{\hbar^2}{\bf m}_{12}^{\rm T}\gamma^{-1}{\bf m}_{12})\Delta \vv q
\nonumber
\\
& -\frac{1}{2}\Delta \vv q^{\rm T} ({\bf m}_{22}^{\rm T}\gamma {\bf m}_{21}+\frac{1}{\hbar^2}{\bf m}_{12}^{\rm T}\gamma^{-1}{\bf m}_{11}) \Delta \vv p\Bigr\}
\end{align}
for the linearized semiclassical (LSC-IVR) result emerges.


The expression in (\ref{eq:R1inter}), however, can be further simplified by noting that
\be
I=\int \frac{{\rm d}^{2N}\Delta z}{(2\pi\hbar)^N}|C(\bar{\vv z},t)|^2
\exp\Bigl\{-\Delta \vv z^{\rm T}{\bf A}\Delta \vv z\Bigr\}
=\frac{|C(\bar{\vv z},t)|^2}{(2\pi\hbar)^N}\left(\frac{\pi^{2N}}{{\rm det}{\bf A}}\right)^{1/2}
\ee
holds for the integral over the difference coordinate, with 
\begin{align}
|C(\bar{\vv z},t)|^2=&\sqrt{{\rm det}
\left(\frac{1}{2}[{\bf m}_{11}+\gamma{\bf m}_{22}\gamma^{-1}+i\hbar\gamma{\bf m}_{21}+\frac{1}{i\hbar}{\bf m}_{12}\gamma^{-1}]\right)}
\nonumber
\\
&\sqrt{{\rm det}\left(\frac{1}{2}[{\bf m}_{11}+\gamma{\bf m}_{22}\gamma^{-1}-i\hbar\gamma{\bf m}_{21}-\frac{1}{i\hbar}{\bf m}_{12}\gamma^{-1}]\right)}
\end{align}
and 
\be
{\bf A}=\frac{1}{4\hbar^2}
\left(\begin{array}{cc}
\gamma^{-1}+{\bf m}_{11}^{\rm T}\gamma^{-1} {\bf m}_{11}+{\bf m}_{21}^{\rm T}\hbar^2\gamma{\bf m}_{21}
&
{\bf m}_{21}^{\rm T}\hbar^2\gamma {\bf m}_{22}+{\bf m}_{11}^{\rm T}\gamma^{-1}{\bf m}_{12})
\\
{\bf m}_{22}^{\rm T}\hbar^2\gamma {\bf m}_{21}+{\bf m}_{12}^{\rm T}\gamma^{-1}{\bf m}_{11}
&
\hbar^2\gamma +{\bf m}_{22}^{\rm T}\hbar^2\gamma {\bf m}_{22}+{\bf m}_{12}^{\rm T}\gamma^{-1}{\bf m}_{12}
\end{array}\right).
\ee
To proceed, we work with the definitions
\bea
\label{eq:r}
{\bf r}&\equiv&{\bf m}_{21}^{\rm T}\gamma+\frac{i}{\hbar}{\bf m}_{11}^{\rm T}
\\
\label{eq:s}
{\bf s}&\equiv&{\bf m}_{22}^{\rm T}\gamma+\frac{i}{\hbar}{\bf m}_{12}^{\rm T}
\eea
and get
\be
|C(\bar{\vv z},t)|^2=\sqrt{{\rm det}\left(\frac{1}{2}[i\hbar {\bf r}^\dagger+{\bf s}^\dagger\gamma^{-1}]\right)
{\rm det}\left(\frac{1}{2}[-i\hbar {\bf r}+\gamma^{-1}{\bf s}]^{\rm T}\right)}
\ee
as well as
\be
{\bf A}=\frac{1}{4}
\left(\begin{array}{cc}
(\gamma^{-1}/\hbar^2+{\bf r}\gamma^{-1}{\bf r}^\dagger)
&({\bf r}\gamma^{-1}{\bf s}^\dagger -i/\hbar)
\\
({\bf s}\gamma^{-1}{\bf r}^\dagger +i/\hbar)
&(\gamma+{\bf s}\gamma^{-1}{\bf s}^\dagger)
\end{array}\right),
\ee
where we have used the relations \cite{jcp06}
\bea
{\bf m}_{22}^T{\bf m}_{11}-{\bf m}_{12}^T{\bf m}_{21}&=&{\bf 1}
\\
{\bf m}_{11}^T{\bf m}_{21}-{\bf m}_{21}^T{\bf m}_{11}&=&{\bf 0}
\\
{\bf m}_{22}^T{\bf m}_{12}-{\bf m}_{12}^T{\bf m}_{22}&=&{\bf 0},
\eea
valid for the sub matrices of the monodromy matrix and where the superscript $\dagger$ indicates the (hermitian) adjunct of the matrix.

The determinant of the block matrix is given by
\be
{\rm det}{\bf A}={\rm det}
\left(\begin{array}{cc}
{\bf a}_{11}&{\bf a}_{12}
\\
{\bf a}_{21}&{\bf a}_{22}
\end{array}\right)
=\det({\bf a}_{11}{\bf a}_{22}-{\bf a}_{11}{\bf a}_{12}^{\rm T}{\bf a}_{11}^{-1}{\bf a}_{12}).
\label{eq:block}
\ee
After a bit of algebra we can manipulate the difference of block matrix products into the helpful
intermediate form 
\be
{\bf a}_{11}{\bf a}_{22}-{\bf a}_{11}{\bf a}_{12}^{\rm T}{\bf a}_{11}^{-1}{\bf a}_{12}
=\frac{1}{16}[{\bf r}\gamma^{-1}+\frac{i}{\hbar}{\bf a}_{11}{\bf s}\gamma^{-1}{\bf r}^\dagger{\bf a}_{11}^{-1}({\bf r}^\dagger\gamma)^{-1}]
({\bf r}^\dagger\gamma-\frac{i}{\hbar}{\bf s}^\dagger),
\ee
the determinant of which is, due to $\det({\bf ab})=\det{\bf a} \det{\bf b}$, given by
\be
{\rm det}{\bf A}=\left(\frac{1}{4\hbar^2}\right)^N
{\rm det}\left(\frac{1}{2}[i\hbar {\bf r}^\dagger+{\bf s}^\dagger\gamma^{-1}]\right)
{\rm det}\left(\frac{1}{2}[-i\hbar {\bf r}+\gamma^{-1}{\bf s}]\right).
\ee
Now because of $\det{\bf a}=\det{\bf a}^{\rm T}$ this cancels the determinants from the preexponential factor as well as all constants and  
we get
\be
I=1,
\ee
a result that, albeit along different lines, has been proven before \cite{HC99,NEL03}. 

The final result for the linear IR correlation function in the linearized semiclassical approximation therefore is
\begin{align}
R^{(1)}_{\rm lsc}(t)=\frac{\beta}{m (2\pi\hbar)^{N} Q}\int {\rm d}^{2N}\bar{z}
\exp\{-\beta H(\bar{\vv z})\}\bar{\vv p}^{\rm T}\bar{\vv q}(t),
\label{eq:lscivr}
\end{align}
with
\be
Q=\frac{1}{(2\pi\hbar)^{N}}\int {\rm d}^{2N}\bar{z}\exp\{-\beta H(\bar{\vv z})\}.
\ee
The quantity $\hbar$ only enters in the prefactor of the final expression in the same manner as in classical statistical mechanics and does not
appear together with any dynamical quantities any more. Therefore this is a classical result, sometimes also called the classical Wigner result.
We note that this final result can also be proven along different lines. In \cite{SG03}, e. g.,  the derivation started directly from
the path integral, without invoking an intermediate semiclassical approximation. In  \cite{HC99}, however, it was shown that this result can 
be gained by starting from a double phase space integral with HK propagators and, instead of the linearization recapitulated here, by doing 
a {\it stationary phase} integration in the difference variable
in the high temperature limit (for low temperatures, the stationary phase condition $\vv z_1=\vv z_2$ is only approximately fulfilled!).

Furthermore, if the two time evolution operators in the original expression (\ref{eq:R1})  are due to different Hamiltonians and
both position operators are replaced by unit operators, then calculus analogous to the
one reviewed here leads to the an expression still involving a phase factor, which is called dephasing representation of fidelity decay \cite{Van06}.
Finally, we note that the calculations performed in this appendix become trivial in the case of $N=1$, i. e., for a single degree of freedom,
because then the sub-block matrices in (\ref{eq:block}) become numbers and do commute!

\section*{References}

\end{document}